\documentclass[aps,pra,reprint,amsmath,amssymb,amsfonts,showkeys,showpacs,preprintnumbers]{revtex4-1}
\usepackage{bm,hyperref,microtype,xcolor}

\begin{document}

\title{Leading-order behavior of the correlation energy in the uniform electron gas}

\author{Pierre-Fran\c{c}ois Loos}
\email{loos@rsc.anu.edu.au}
\affiliation{Research School of Chemistry, 
Australian National University, Canberra, ACT 0200, Australia}
\author{Peter M. W. Gill}
\thanks{Corresponding author}
\email{peter.gill@anu.edu.au}
\affiliation{Research School of Chemistry, 
Australian National University, Canberra, ACT 0200, Australia}
\date{\today}

\begin{abstract}
We show that, in the high-density limit, restricted M{\o}ller-Plesset (RMP) perturbation theory yields $E_{\text{RMP}}^{(2)} = \pi^{-2}(1-\ln 2) \ln r_s + O\left(r_s^0\right)$ for the correlation energy per electron in the uniform electron gas, where $r_s$ is the Seitz radius.
This contradicts an earlier derivation which yielded $E_{\text{RMP}}^{(2)} = O(\ln\left|\ln r_s\right|)$.  The reason for the discrepancy is explained.
\end{abstract}

\maketitle

We consider a paramagnetic system of $N$ interacting electrons confined in a cubic box, with edges of length $L$ and volume $\Omega=L^3$.
We also assume a uniform charge density background of density equal in magnitude, but opposite in sign, to the average electron density $\rho = N/\Omega$.  In the thermodynamic limit, both $N$ and $\Omega$ tend to infinity in such a way that the system becomes homogeneous with a uniform density $\rho$, related to the Seitz radius by the relation $r_s = \left(4\pi\rho/3\right)^{-1/3}$, and is often called jellium \cite{Vignale, ParrYang}.

It is convenient to consider a reduced Hamiltonian (\textit{i.e.}~one that is scaled by the number of electrons) and, in atomic units, this is
\begin{equation}
\label{H}
	\Hat{H} =  N^{-1} \left[ \Hat{T} + \Hat{H}_{\text{e-e}} + \Hat{H}_{\text{e-b}} + \Hat{H}_{\text{b-b}} \right],
\end{equation}
where the operator
\begin{equation}
	\Hat{T} = 
	- \frac{1}{2} \sum_{i=1}^{N} \nabla_i^2
\end{equation}
corresponds to the kinetic energy of the electrons, and
\begin{align}
	\Hat{H}_{\text{e-e}} & = 
	\sum_{i<j}^{N} \frac{1}{\left|\bm{r}_i-\bm{r}_j\right|},
	\\
	\Hat{H}_{\text{e-b}} & = 
	-\rho \sum_{i=1}^{N} \int 
	\frac{d\bm{R}}{\left|\bm{r}_i-\bm{R}\right|},
	\\
	\Hat{H}_{\text{b-b}} & = 
	\frac{\rho^2}{2} \iint 
	\frac{d\bm{R}_1 d\bm{R}_2}{\left|\bm{R}_1-\bm{R}_2\right|}
\end{align}
represent the electron-electron, electron-background and background-background interactions, respectively.
\footnote{In the thermodynamic limit, both $\Hat{H}_{\text{b-b}}$ and $\Hat{H}_{\text{e-b}}$ diverge.
However, the divergence is cancelled by a term from $\Hat{H}_{\text{e-e}}$. 
See Ref. \onlinecite{Vignale} for more details.}
The term $\Hat{H}_{\text{b-b}}$ is a known constant \cite{Jellium05} and may be ignored.

In perturbation theory\cite{Helgaker}, we introduce a partition
\begin{equation}
	\Hat{H} = \Hat{H}^{(0)} + \Hat{V},
\end{equation}
where the perturbation $\Hat{V}$ is assumed small (in some sense) compared to the zeroth-order Hamiltonian $\Hat{H}^{(0)}$.  This yields an expansion of the (reduced) energy
\begin{equation}
\label{E}
	E = E^{(0)} + E^{(1)} + E^{(2)} + \ldots.
\end{equation}
The zeroth-, first- and second-order energies are given by
\begin{align}
	E^{(0,\ell)} & = \left< \Psi^{(0,\ell)} \left| \Hat{H}^{(0)} \right| \Psi^{(0,\ell)} \right>,
	\label{E0}
	\\
	E^{(1)} & = \left< \Psi^{(0)} \left| \Hat{V} \right| \Psi^{(0)} \right>,
	\label{E1}
	\\
	E^{(2)} & = \sum_{\ell=1}^{\infty} 
	\frac{\left< \Psi^{(0,\ell)} \left| \Hat{V} \right| \Psi^{(0)} \right>^2}
	{E^{(0,\ell)} - E^{(0)}},
	\label{E2}
\end{align}
where $E^{(0,0)} \equiv E^{(0)}$ and $\Psi^{(0,0)} \equiv \Psi^{(0)}$,
and the zeroth-order ground state ($\ell=0$) and excited states ($\ell>0$) wave functions satisfy
\begin{equation}
	\Hat{H}^{(0)} \Psi^{(0,\ell)} = E^{(0,\ell)} \Psi^{(0,\ell)}.
\end{equation}
There are many ways to partition $\Hat{H}$ but not all are equally effective.  In this Brief Report, we will consider three: the non-interacting (NI), restricted M{\o}ller-Plesset \cite{Moller34} (RMP), and unrestricted M{\o}ller-Plesset (UMP) partitions.

If we adopt the NI partition, we have
\begin{align}
	\Hat{H}_{\text{NI}}^{(0)} & = \Hat{T},
	&
	\Hat{V}_{\text{NI}} & = \Hat{H}_{\text{e-e}} + \Hat{H}_{\text{e-b}}.
\end{align}
The zeroth-order wave functions $\Psi_{\text{NI}}^{(0,\ell)}$ are Slater determinants of plane-wave orbitals
\begin{equation}
\label{eigenvec}
	\psi_{\bm{k}}\left(\bm{r}\right) = \frac{1}{\Omega} \exp\left(i\,\bm{k}\cdot\bm{r}\right),
\end{equation}
with orbital energies
\begin{equation}
\label{eigenval-H}
	\epsilon_{\text{NI}}(k) = \frac{k^2}{2}.
\end{equation}
The $\ell$-th excited determinant $\Psi_{\text{NI}}^{(0,\ell)}$ has the energy
\begin{equation}
	E_{\text{NI}}^{(0,\ell)} = \frac{1}{N} \sum_{\bm{k}}^{\text{occ}} \epsilon_{\text{NI}}(k),
\end{equation}
where the sum over $\bm{k}$ takes into account all the plane waves used to build $\Psi_{\text{NI}}^{(0,\ell)}$, {\em i.e.} all the occupied orbitals in the state $\ell$.
For the special case $\ell=0$, all the orbitals up to the Fermi level are occupied.

Introducing $\alpha=\left(9\pi/4\right)^{1/3}$, one finds \cite{Fermi26, Thomas27, Dirac30} that
\begin{align}
\label{E-0-1-NI}
	E_{\text{NI}}^{(0)} & = \frac{3}{10} \frac{\alpha^2}{r_s^2},	&
	E_{\text{NI}}^{(1)} & = - \frac{3}{4\pi} \frac{\alpha}{r_s},
\end{align}
which are the kinetic and exchange energies, respectively. 
\footnote{The Coulomb part in $E_{\text{NI}}^{(1)}$ is exactly cancelled by the positive uniform background {\em via} the term $\Hat{H}_{\text{e-b}}$.}

Unfortunately, although the correlation energy \cite{Wigner34}
\begin{equation}
	E_{\text{c}} = E - E^{(0)} - E^{(1)}
\end{equation}
of jellium is known \cite{Ceperley80} to be finite for any $r_s > 0$, the second-order energy Eq.~\eqref{E2} is infinite.  However, the leading-order contribution can be extracted from Eq.~\eqref{E2} and, henceforth, we will use $E^{(2)}$ to refer to that contribution.

After transforming into momentum space and scaling the momenta by the wave vector $k_{\text{F}}=\alpha/r_s$ so that the Fermi sphere has unit radius, one finds \cite{Macke50, GellMann57} that $E_{\text{NI}}^{(2)}$ consists of a direct (``ring-diagram'') contribution 
\begin{equation}
\label{E2r}
	E_{\text{NI}}^{(2,\text{a})} = - \frac{3}{16\pi^5} \iiint \frac{d\bm{q}\,d\bm{k}_1\,d\bm{k}_2}{q^4 \Delta\epsilon_{\text{NI}}},
\end{equation}
and an exchange contribution
\begin{equation}
\label{E2x}
	E_{\text{NI}}^{(2,\text{b})} = \frac{3}{32\pi^5} \iiint
		\frac{d\bm{q}\,d\bm{k}_1\,d\bm{k}_2}{q^2|\bm{q}+\bm{k}_1-\bm{k}_2|^2 \Delta\epsilon_{\text{NI}}}.
\end{equation}
In these integrals, the excitation vector $\bm{q}$ has the domain
\begin{equation}
\label{domain-q}
	\beta < |\bm{q}| < \infty,
\end{equation}
where $\beta \propto \sqrt{r_s}$ \cite{Bohm53}, and the occupied-orbital vectors $\bm{k}_1$ and $\bm{k}_2$ have domains 
\begin{align}
	&	|\bm{k}_1| < 1,	&	|\bm{k}_1+\bm{q}| > 1,	\label{domain1}	\\
	&	|\bm{k}_2| < 1,	&	|\bm{k}_2-\bm{q}| > 1,	\label{domain2}
\end{align}
The lower bound for $q$ in Eq.~\eqref{domain-q} is due to the screening effect of the Coulomb field by the collective electron motions, and can be derived using the plasma theory of the free-electron gas \cite{Bohm51, Pines52, Bohm53, Pines53, Nozieres58}.
The orbital energy difference is
\begin{multline}
	\Delta\epsilon_{\text{NI}} 
	= \epsilon_{\text{NI}}\left(\left|\bm{k}_1+\bm{q}\right|\right)
	+ \epsilon_{\text{NI}}\left(\left|\bm{k}_2-\bm{q}\right|\right)
	\\
	- \epsilon_{\text{NI}}(k_1) - \epsilon_{\text{NI}}(k_2).
\end{multline}
The exchange term $E_{\text{NI}}^{(2,\text{b})}$ is finite \cite{Onsager66} and, for small $r_s$, is dominated by the ring-diagram term 
\begin{equation}
	E_{\text{NI}}^{(2,\text{a})} = \frac{1 - \ln 2}{\pi^2} \ln r_s + O(r_s^{0}),
\end{equation}
which Macke showed \cite{Macke50} to depend logarithmically on $r_s$.  One may wonder, however, whether this logarithmic term arises when the Hamiltonian is partitioned differently \cite{Handler88}.

If we adopt the RMP partition \cite{Szabo}, we have
\begin{align}
	\Hat{H}_{\text{RMP}}^{(0)} & = \sum_{i=1}^{N} \Hat{F}\left(\bm{r}_i\right),
	& 
	\Hat{V}_{\text{RMP}} & = \Hat{H} - \sum_{i=1}^{N} \Hat{F}\left(\bm{r}_i\right),
\end{align}
where the Fock operator defined by
\begin{multline}
\label{F}
	\Hat{F}(\bm{r}_1) \psi_{\bm{k}_1}(\bm{r}_1)
	= - \frac{1}{2} \nabla_1^2 \psi_{\bm{k}_1}(\bm{r}_1) 
	\\
	+ \sum_{\bm{k}_2}^{\text{occ}} \psi_{\bm{k}_2}(\bm{r}_1) 
	\int
	\frac{\psi_{\bm{k}_2}^{*}(\bm{r}_2) \psi_{\bm{k}_1}(\bm{r}_2)}{|\bm{r}_1-\bm{r}_2|} d\bm{r}_2 
\end{multline}
includes kinetic and exchange terms but not Hartree terms because of their cancelation by the $\Hat{H}_{\text{e-b}}$ term.

The RMP zeroth-order wave functions $\Psi_{\text{RMP}}^{(0,\ell)}$ are again determinants of plane-wave orbitals \eqref{eigenvec}, but the orbital energies are now different and it can be shown \cite{Raimes1, Raimes2} that 
\begin{equation}
\label{eigenval-HF}
	\epsilon_{\text{RMP}}(k) 
	= \epsilon_{\text{NI}}(k) 
	- \frac{r_s}{\alpha\pi} \epsilon_{\text{X}}(k),
\end{equation}
The additional term
\begin{equation}
	\epsilon_{\text{X}}(k) = 1 + \frac{1-k^2}{2k} \ln \left|\frac{1+k}{1-k}\right|
\end{equation}
arises from the exchange terms in Eq.~\eqref{F}.  Thus,
\begin{equation}
	\Psi_{\text{RMP}}^{(0,\ell)} = \Psi_{\text{NI}}^{(0,\ell)},
\end{equation}
but
\begin{equation}
	E_{\text{RMP}}^{(0,\ell)} 
	= \frac{1}{N} \sum_{\bm{k}}^{\text{occ}} \epsilon_{\text{RMP}}(k)
	\neq E_{\text{NI}}^{(0,\ell)}.
\end{equation}

The zeroth- and first-order energies are now given by
\begin{align}
\label{E-0-1-MP}
	E_{\text{RMP}}^{(0)} & = 
	\frac{3}{10} \frac{\alpha^2}{r_s^2} 
	- \frac{3}{2\pi} \frac{\alpha}{r_s},
	&
	E_{\text{RMP}}^{(1)} & = 
	\frac{3}{4\pi} \frac{\alpha}{r_s},
\end{align}
and comparing Eqs \eqref{E-0-1-NI} and \eqref{E-0-1-MP} reveals the important relation
\begin{equation}
	E_{\text{NI}}^{(0)} + E_{\text{NI}}^{(1)} = E_{\text{RMP}}^{(0)} + E_{\text{RMP}}^{(1)} = E_{\text{RHF}},
\end{equation}
where $E_{\text{RHF}}$ is the reduced RHF energy.

The ring-diagram contribution to $E_\text{RMP}^{(2)}$ is
\begin{equation}
\label{E2r-MP}
	E_{\text{RMP}}^{(2,\text{a})} = - \frac{3}{16\pi^5} \iiint \frac{d\bm{q}\,d\bm{k}_1\,d\bm{k}_2}{q^4 \Delta\epsilon_{\text{RMP}}},
\end{equation}
which differs from Eq.~\eqref{E2r} only by the denominator
\begin{equation}
	\Delta\epsilon_{\text{RMP}} = \Delta\epsilon_{\text{NI}} - \frac{r_s}{\alpha\pi}\Delta\epsilon_{\text{X}},
\end{equation}
where
\begin{multline}
	\Delta\epsilon_{\text{X}} 
	= \epsilon_{\text{X}}\left(\left|\bm{k}_1+\bm{q}\right|\right)
	+ \epsilon_{\text{X}}\left(\left|\bm{k}_2-\bm{q}\right|\right)
	\\
	- \epsilon_{\text{X}}(k_1) - \epsilon_{\text{X}}(k_2).
\end{multline}
The behavior of $E_{\text{RMP}}^{(2,\text{a})}$ is dominated \cite{Raimes2} by contributions in the neighborhood of the Fermi sphere ({\em i.e.}~$q \approx 0$).  On the domains \eqref{domain1} and \eqref{domain2}, we have
\begin{equation}
	\epsilon_{\text{RMP}}(k) = \frac{k^2}{2} - \frac{r_s}{\alpha\pi}\left[1+\frac{1-k^2}{2k} \ln \frac{1+k}{1-k}\right],
\end{equation}
and
\begin{multline}
	\epsilon_{\text{RMP}}\left(\left|\bm{k}+\bm{q}\right|\right) = \frac{\left|\bm{k}+\bm{q}\right|^2}{2}	
	\\
	- \frac{r_s}{\alpha\pi} \left[1+\frac{1-|\bm{k}+\bm{q}|^2}{2\left|\bm{k}+\bm{q}\right|} 
	\ln \frac{\left|\bm{k}+\bm{q}\right|+1}{\left|\bm{k}+\bm{q}\right|-1}\right].
\end{multline}
Therefore, we have
\begin{equation}
\label{Deps}
	\Delta\epsilon_{\text{RMP}} \approx u + v - \frac{r_s}{\alpha\pi} \left(u \ln \frac{u}{2} + v \ln \frac{v}{2}\right),
\end{equation}
where we have introduced
\begin{align}
	u & = \frac{\bm{k}_1 \cdot \bm{q}}{k_1},
	& 
	v & = - \frac{\bm{k}_2 \cdot \bm{q}}{k_2}.
\end{align}
Substituting \eqref{Deps} into \eqref{E2r-MP} and using the relations 
\begin{gather}
	d\bm{k}_1 = 2\pi k_1^2 \sin\theta\,d\theta\,dk_1 \approx \frac{2\pi}{q}\,du\,dk_1,
	\\
	\left|\bm{k}_1 + \bm{q}\right|>1 \Rightarrow 1 - u \le k_1 \le 1,
\end{gather}
(with similar expressions for $k_2$) then yields
\begin{widetext}
\begin{equation}
\label{E2r-3}
\begin{split}
	E_{\text{RMP}}^{(2,\text{a})}
		&	\approx - \frac{3}{\pi^2} \int_{\beta}^1 \frac{dq}{q^4} \int_0^q du \int_{1-u}^1 dk_1 \int_0^q dv \int_{1-v}^1 dk_2
		\frac{1}{u+v - \frac{r_s}{\alpha\pi} \left(u \ln \frac{u}{2} + v \ln \frac{v}{2}\right)}				
		\\
		&	= - \frac{3}{\pi^2} \int_{\beta}^1 \frac{dq}{q^4} \int_0^q du \int_0^q dv	
		\frac{u\,v}{u+v - \frac{r_s}{\alpha\pi} \left(u \ln \frac{u}{2} + v \ln \frac{v}{2}\right)}.
\end{split}
\end{equation}
Since the most important contribution comes from small $q$, we have set the upper bound of the integral \eqref{E2r-3} to a convenient value of unity.  Expanding for small $r_s$ and integrating over $q$ yields
\begin{equation}
\label{E2r-final}
\begin{split}
	E_{\text{RMP}}^{(2,\text{a})}	& \approx - \frac{3}{\pi^2} \int_{\beta}^1 \frac{dq}{q^4} \int_0^q du \int_0^q dv
					\frac{u\,v}{u+v} \left[1+ \frac{r_s}{\alpha\pi}\frac{u \ln \frac{u}{2} + v \ln \frac{v}{2}}{u+v}\right]
					= \frac{1-\ln 2}{\pi^2} \ln r_s + O(r_s^0),
\end{split}
\end{equation}
\end{widetext}
which is identical, in the high-density ({\em i.e.} small-$r_s$) limit, to $E_{\text{NI}}^{(2,\text{a})}$.  
The present result can also be obtained from \eqref{E2r-3} by first switching to polar coordinates ($u=r \cos \theta$ and $v=r \sin \theta$), integrating over the radial part,  carefully taking the $r_s\to0$ limit, and finally performing the remaining angular integration.  The latter derivation rigorously justifies the small-$r_s$ expansion  \footnote{We thank the referee for providing this alternative derivation.}.

In a similar investigation more than 20 years ago \cite{Handler88}, Handler claimed to show that
\begin{equation}
\label{E2r-HF-Handler}
	E_{\text{RMP}}^{(2,\text{a})} = O(\ln \left| \ln r_s \right|).
\end{equation}
This claim, which implies that $E_{\text{RMP}}^{(2)}$ grows more slowly with $r_s$ than $E_\text{NI}^{(2)}$, obviously disagrees with our result in Eq.~\eqref{E2r-final}.  However, in his analog of Eq.~\eqref{E2r-3}, Handler drops the $u+v$ term and ignores the $r_s/\alpha\pi$ factor \footnote{Although it does not seem to alter Handler's derivation, the expressions reported in Ref. \onlinecite{Handler88} for $\epsilon_{\text{RMP}}\left(\left|\bm{k}_1+\bm{q}\right|\right)$ and $\epsilon_{\text{RMP}}(k_1)$ are incorrect.}.  The fact that $\beta \propto \sqrt{r_s}$ means that Handler's neglect of the $u+v$ term is incorrect.

It may be surprising that $E_\text{RMP}^{(2)}$ is the same as $E_\text{NI}^{(2)}$, because $\Hat{H}_{\text{RMP}}^{(0)}$ seems a better starting point than $\Hat{H}_{\text{NI}}^{(0)}$.  However, this is not the first time that the RHF treatment of jellium has been disappointing.  For example, the RHF bandwidth, $\epsilon(1)-\epsilon(0)$, is greater than the NI bandwidth, which disagrees with experiments on simple metals, where a small reduction is observed \cite{Jensen85, Lyo88}.  Moreover, the logarithmic dependence of the eigenvalues \eqref{eigenval-HF} leads to a divergent derivative of $\epsilon_{\text{RMP}}(k)$ at the surface of the Fermi sphere ($k = 1$) and this leads to incorrect dependence of the electronic specific heat on temperature.  Experimentally, a linear dependence with a prefactor close to the NI value is observed \cite{Bardeen36, Bohm50}.

One may hope that a different, and superior, perturbation series can be obtained by adopting the UMP partition, that is, by using the UHF wavefunction of jellium as the starting point.  After all, as Overhauser showed long ago \cite{Overhauser59, Overhauser62}, the RHF solution of jellium is unstable with respect to a lower-energy UHF solution, for all $r_s$ \cite{Zhang2008}.  However, we expect that there will be serious issues with the convergence of the UMP perturbation series \cite{Deceptive86, SlowUMP88} and we have not considered this alternative in detail.

In conclusion, we have shown that the correlation energy $E_\text{RMP}^{(2)}$ from RMP perturbation theory, {\em i.e.}~using a RHF starting point, is the same as the $E_\text{NI}^{(2)}$ from conventional NI perturbation theory.  Although it is nearly impossible to test experimentally this result,  this corrects an earlier study which claimed that $E_{\text{RMP}}^{(2)}$ is sub-logarithmic.

The authors thank Joshua Hollett for many valuable discussions, and the referee for his careful analysis of the present manuscript.  P.M.W.G. thanks the NCI National Facility for a generous grant of supercomputer time and the Australian Research Council (Grants DP0984806 and DP1094170) for funding.

\end{document}